# State Orthogonality, Boson Bunching Parameter and Bosonic Enhancement Factor

Avi Marchewka[1] and Er'el Granot[2]

1) 8 Galei Tchelet St. Herzliya, Israel; *email:avi.marchewka@gmail.com*
2) Department of Electrical and Electronics Engineering, Ariel University, 40700 Ariel, Israel; *email: erelgranot@gmail.com*

**Abstract**

It is emphasized that the bunching parameter $\beta \equiv p_B / p_D$, i.e. the ratio between the probability to measure two bosons and two distinguishable particles at the same state, is a constant of motion and depends only on the overlap between the initial wavefunctions. This ratio is equal to $\beta = 2/(1+I^2)$, where $I$ is the overlap integral between the initial wavefunctions. That is, only when the initial wavefunctions are orthogonal this ratio is equal to 2, however, this bunching ratio can be reduced to 1, when the two wavefunctions are identical. This simple equation explains the experimental evidences of a beam splitter. A straightforward conclusion is that by measuring the *local* bunching parameter $\beta$ (at any point in space and time) it is possible to evaluate a *global* parameter $I$ (the overlap between the initial wavefunctions). The bunching parameter is then generalized to arbitrary number of particles, and in an analogy to the two-particles scenario, the well-known bosonic enhancement appears only when all states are orthogonal.

## 1.Introduction

Bosons bunching (BB) is the tendency of bosons to bunch together with respect to distinguishable particles. There are many applications and experimental evidences as well as theoretical ones, which indeed validate this property of bosons[1-13], where the term bosons is used for particles whose eigenstate is symmetric under particle exchange in the relevant subspace. On the face of it, it seems that the theoretical basis for bunching is sound, however, recently, it has been shown that in some scenarios bosons can behave like distinguishable particles and can even antibunch like fermions[14]. It has also been shown that in some scenarios bosons behave as if they repel each other more than fermions [15].

There is a small distinction between the different descriptions of bunching, since there are small, albeit important, nuances in the measurements procedures (see a discussion in [14], which is based on [16] and [17]). In these previous works, bosons non-bunching and bosons anti-bunching were demonstrated for Feynman's description of bunching [16,18].

In this paper we focus on what seems to be the basic and common description of bunching: bunching occurs when the bunching parameter $\beta \equiv p_B / p_D$, i.e., the ratio between the probabilities to find two bosons ($p_B$) and two distinguishable particles ($p_D$) *at the same state*, is larger than 1 (see below Eq.(3)).

The bunching parameter can be measured by performing the same experiment twice: for bosons and for distinguishable particles. Both experiments are repeated for the same amount of events, and in both experiments the number of events, in which both particles were measured in the same state is recorded. The ratio between these numbers is an evaluation of the bunching parameter of this state. In many cases it is unnecessary to measure the distinguishable particle part of the experiment, however, if it is required then the simplest way to distinguish between the particle is by adding a marker on one of them, i.e., to die them differently. This is usually done by controlling the polarization (in case of photons) or spins (in case of massive particles) of the particles differently (see, for example Ref.[12]). When the particles spins/polarizations are unknown then they can be regarded as bosons, but if one has spin up (linear polarization in the x-direction) and the other has spin down (linear polarization in the y-direction), then they can be regarded as distinguishable. The experimentalist can use the same setup and switch between bosons experiment to distinguishable particles one by activating a spin/polarization rotaor on one of the particles.

Another option to simulate distinguishable particles experiment without changing the experimental setup is to take advantage of the fact that the probability to measure two distinguishable particles at the same state is simply the product of the single particle probabilities to reach the same state (see Eq.(2) below). Therefore, the bunching parameter can be evaluated by repeating the same experiment thrice: in the first experiment the two sources (states, $|\psi_1\rangle$ and $|\psi_2\rangle$) are present and the bosons probability to reach the same state $|m\rangle$ is detected $p_B$. In the second experiment one of the sources (one of the states, say $|\psi_1\rangle$) is omitted, and then the detector measures the single particle probability $|\langle m|\psi_2\rangle|^2$. Similarly, in the third experiement the second source (state) is omitted and the detector measures the probability $|\langle m|\psi_1\rangle|^2$. The ratio between $p_B$ and the product $|\langle m|\psi_1\rangle|^2 |\langle m|\psi_2\rangle|^2$ is the bunching parameter.

A great amount of important research took place in many-body experiments in general, and in quantum optics in particular, to utilize first and second order correlation functions to quantify a source correlation[19-20]. There is some similarity between the bunching parameter and the second order correlation function; however, the terms are not identical, and it will be explained why the bunching parameter is more useful to our purposes.

It should be stressed that in this paper we use the term bosons to describe "spinless bosons" and the term fermions to describe "spinless fermions", i.e., we focus on the spatial part of the wavefunction. Therefore, bosons/fermions would have symmetric/antisymmetric spatial joint wavefunction. However, as will be shown at the end of the paper, the results of this paper can be applied to any degree of freedom including spin.

## 2. Orthogonal States

In textbooks (see, for example [17]) one can find the following reasoning. Let $|\psi_1\rangle$ and $|\psi_2\rangle$ be the two initial *orthonormal* states of the two bosons (i.e., we know that one of them occupies state $|\psi_1\rangle$ and the second occupies state $|\psi_2\rangle$), and let $|m\rangle$ and $|p\rangle$ be the final states (i.e one particle was measured at $|m\rangle$ and the other at $|p\rangle$) then the probability to measure them at these states is (according to Eqs.D-18 and D-23 of Ref.[17] )

$$p_B = \begin{cases} |\langle m|\psi_1\rangle\langle p|\psi_2\rangle + \langle p|\psi_1\rangle\langle m|\psi_2\rangle|^2 & m \neq p \\ 2|\langle m|\psi_1\rangle\langle m|\psi_2\rangle|^2 & m = p \end{cases} \tag{1}$$

while the probability to measure two distinguishable particles at the same states ($|m\rangle$ and $|p\rangle$) is[17]

$$p_D = \begin{cases} |\langle m|\psi_1\rangle\langle p|\psi_2\rangle|^2 + |\langle p|\psi_1\rangle\langle m|\psi_2\rangle|^2 & m \neq p \\ |\langle m|\psi_1\rangle\langle m|\psi_2\rangle|^2 & m = p \end{cases}. \tag{2}$$

Hence, the probability to measure the two bosons at the *same* state, i.e., $m = p$, is exactly twice the probability to measure the two distinguishable particles there, and therefore the bunching parameter satisfies

$$\beta(m=p) = \frac{p_B(m=p)}{p_D(m=p)} = 2. \tag{3}$$

A well-known example, which illustrates this result, is the beam-splitter.

If, as is often shown, $|\psi_1^{in}\rangle = |u\rangle = \begin{pmatrix} 1 \\ 0 \end{pmatrix}$, stands for particle enters the upper input of the splitter, and $|\psi_2^{in}\rangle = |d\rangle = \begin{pmatrix} 0 \\ 1 \end{pmatrix}$ stands for particle enters the lower input of the splitter, then, due to the splitter's transfer matrix

$$U = \frac{1}{\sqrt{2}}\begin{pmatrix} 1 & i \\ i & 1 \end{pmatrix}, \tag{4}$$

the wavefunctions at the splitter's outputs are $|\psi_1^{out}\rangle = |R\rangle \equiv \frac{1}{\sqrt{2}}(|u\rangle + i|d\rangle) = \frac{1}{\sqrt{2}}\begin{pmatrix}1\\i\end{pmatrix}$

and $|\psi_2^{out}\rangle = |L\rangle \equiv \frac{1}{\sqrt{2}}(i|u\rangle + |d\rangle) = \frac{1}{\sqrt{2}}\begin{pmatrix}i\\1\end{pmatrix}$ respectively (see Fig.1), and therefore the joint input state is

$$|\Psi^{in}\rangle = \frac{1}{\sqrt{2}}\left[|1:u;2:d\rangle + |2:d;1:u\rangle\right] \tag{5}$$

while the joint output state is (see Fig.2)

$$|\Psi^{out}\rangle = \frac{i}{\sqrt{2}}\left[|1:u;2:u\rangle + |1:d;2:d\rangle\right]. \tag{6}$$

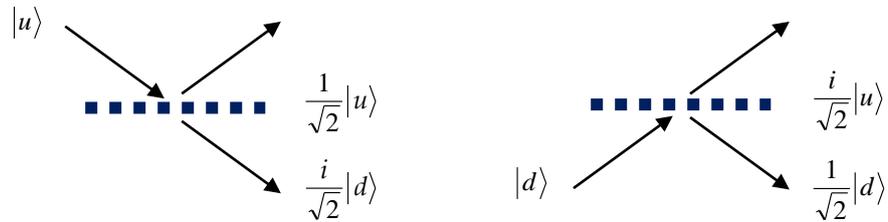

Figure 1: The beam splitter operation on the up (left) and down (right) states.

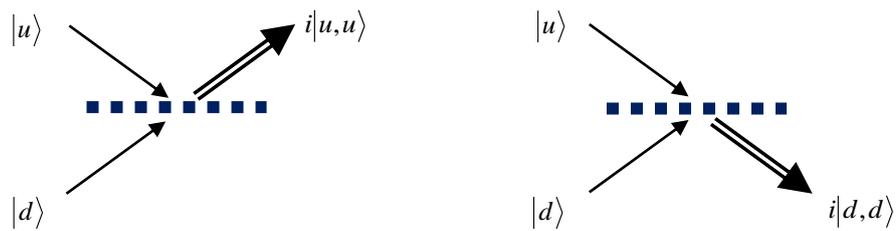

Figure 2: Bosons bunching. If the bosons enter both inputs of the beam splitter they emerge together at the exits.

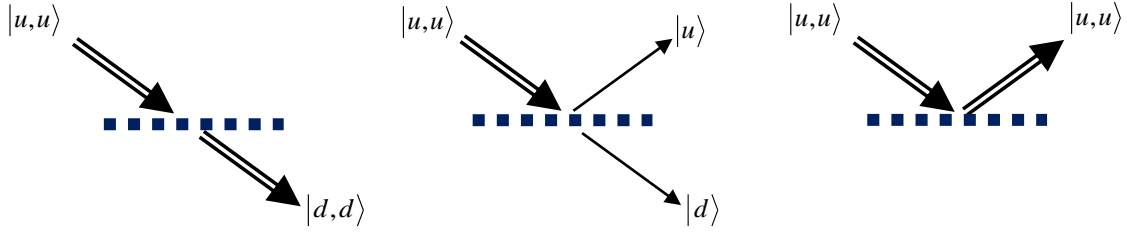

Figure 3: When the bosons enter the splitter from the same input they behave like distinguishable particles – all combinations are possible.

Hence, the probability to measure both particles at the same output (say the upper one) is (see Fig.2)

$$p_B = \left|\langle 1:u;2:u | \Psi^{out}\rangle\right|^2 = \frac{1}{2}, \tag{7}$$

which is exactly twice the probability to measure two distinguishable particles there

$$p_D = \left|\langle u | \psi_1^{out}\rangle\langle u | \psi_2^{out}\rangle\right|^2 = \frac{1}{4}. \tag{8}$$

This result is particularly important since it shows that while distinguishable particles can split (one can go up and one down) with probability 0.5, two bosons must exit together, i.e. both up or both down. Hence, this result apparently emphasizes the bunching property of bosons.

## 3. Identical States

The derivation of the previous section was based on the premise that the two initial states were orthogonal. Not only that the derivation of Eq.(1) was based on the orthogonal premise, but so does the beam splitter initial and final states: $\left|\psi_1^{in}\right\rangle$ and $\left|\psi_2^{in}\right\rangle$ are orthogonal and so do $\left|\psi_1^{out}\right\rangle$ and $\left|\psi_2^{out}\right\rangle$ (which is clear from the unitarity of the splitter).

It can easily be shown that by taking two non-orthogonal initial states the final results are totally different. For example, if the initial states of the particles entering the beam splitter are the same, $\left|\psi_1^{in}\right\rangle = \left|\psi_2^{in}\right\rangle = |u\rangle = \begin{pmatrix} 1 \\ 0 \end{pmatrix}$ then so do the output states

$$\left|\psi_1^{out}\right\rangle = \left|\psi_2^{out}\right\rangle = \frac{1}{\sqrt{2}}(|u\rangle + i|d\rangle) = \frac{1}{\sqrt{2}}\begin{pmatrix} 1 \\ i \end{pmatrix}.$$

The joint input and output states are then

$$\left|\Psi^{in}\right\rangle = |1:u;2:u\rangle \text{ and } \left|\Psi^{out}\right\rangle = |1:R;2:R\rangle \qquad (9)$$

respectively, in which case

$$p_B = \left|\langle 1:u;2:u | \Psi^{out}\rangle\right|^2 = \frac{1}{4}. \qquad (10)$$

Therefore, in this case bosons conduct is exactly similar to the conduct of distinguishable particles (see Fig.3)

$$\beta(m=p) = \frac{p_B(m=p)}{p_D(m=p)} = 1. \qquad (11)$$

Therefore, even from this simple example it is clear that bosons do not always bunch at the output of the beam splitter (as is well known).

## 4. The Generic Two-Particle Scenario and Non-Orthogonality

To show that bunching is indeed an effect of state orthogonality, we solve the generic case.

We begin with a system with a set of discrete eigenstates $|k\rangle$, where $k = 0,1,2,\ldots$. This can be either a finite or an infinite set. Without loss of generality, the single particle initial state can be written as a superposition of these eigenstates:

$$|\psi_1^{in}\rangle = \sum_k \langle k|\psi_1^{in}\rangle |k\rangle, \quad |\psi_2^{in}\rangle = \sum_k \langle k|\psi_2^{in}\rangle |k\rangle \qquad (12)$$

where it is assumed that both states are normalized, i.e., $\sum_k |\langle k|\psi_1^{in}\rangle|^2 = 1$ and $\sum_k |\langle k|\psi_2^{in}\rangle|^2 = 1$. Then, the joint initial state is

$$|\Psi^{in}\rangle = \frac{1}{\sqrt{2N}} \left[ |1:\psi_1; 2:\psi_2\rangle + |2:\psi_1; 1:\psi_2\rangle \right] \qquad (13)$$

and the probability to measure the particles in states $m$ and $p$ is

$$p_B^{in}(m,p) = \begin{cases} |\langle m|\psi_1\rangle\langle p|\psi_2\rangle + \langle p|\psi_1\rangle\langle m|\psi_2\rangle|^2 / N & m \neq p \\ 2|\langle m|\psi_1\rangle\langle m|\psi_2\rangle|^2 / N & m = p \end{cases} \qquad (14)$$

where now the normalization constant is (instead of 1)

$$N \equiv \frac{1}{2} \sum_{m,p} |\langle m|\psi_1\rangle\langle p|\psi_2\rangle + \langle p|\psi_1\rangle\langle m|\psi_2\rangle|^2 = 1 + |I|^2 \qquad (15)$$

and

$$|I| \equiv \left| \sum_m \langle \psi_1|m\rangle\langle m|\psi_2\rangle \right| = |\langle \psi_1|\psi_2\rangle| \qquad (16)$$

is the scalar product between the initial states.

Similarly, if the propagator operator of the system is $U$ then the output (or final) states are

$$|\psi_1^{out}\rangle = \sum_k \langle k|\psi_1^{in}\rangle U|k\rangle \quad \text{and} \quad |\psi_2^{out}\rangle = \sum_k \langle k|\psi_2^{in}\rangle U|k\rangle \qquad (17)$$

and the joint output (or final) state is

$$|\Psi^{out}\rangle = \frac{1}{\sqrt{2N}} \left[ |1:U\psi_1; 2:U\psi_2\rangle + |2:U\psi_1; 1:U\psi_2\rangle \right] \qquad (18)$$

and the probability to measure the bosons in states $m$ and $p$ is

$$p_B^{out}(m,p) = \begin{cases} \left|\langle m|U\psi_1\rangle\langle p|U\psi_2\rangle + \langle p|U\psi_1\rangle\langle m|U\psi_2\rangle\right|^2 / N & m \neq p \\ 2\left|\langle m|U\psi_1\rangle\langle m|U\psi_2\rangle\right|^2 / N & m = p \end{cases}. \quad (19)$$

Due to the unitarity of the operator $U$, there is no difference in the normalization constant, i.e.,

$$N = 1 + |I|^2 \quad (20)$$

since

$$\left|I^{out}\right| \equiv \left|\sum_m \langle\psi_1|U^T|m\rangle\langle m|U|\psi_2\rangle\right| = \left|\langle\psi_1|U^TU|\psi_2\rangle\right| = \left|\langle\psi_1|\psi_2\rangle\right| = \left|I^{in}\right|, \quad (21)$$

which is a real number that can vary between 0 and 1.

Therefore, the probability to find the two bosons in the same state $m$ is:

$$p_B = \frac{2\left|\langle m|U|\psi_1\rangle\langle m|U|\psi_2\rangle\right|^2}{1 + |I|^2}, \quad (22)$$

while the corresponding probability for distinguishable particles is

$$p_D = \left|\langle m|U|\psi_1\rangle\langle m|U|\psi_2\rangle\right|^2. \quad (23)$$

Hence, the ratio again satisfies

$$1 \leq \beta = \frac{2}{1 + |I|^2} \leq 2 \quad (24)$$

and it is totally independent of the Hamiltonian of the system.

Hence, the only thing that determines the bunching properties of the particles at a given system is the scalar product (or orthogonality) between the two initial states.

It is therefore clear that two initially orthogonal states, i.e., $I=0$, causes bunching, i.e., $\beta = p_B/p_D = 2$, while identical initial states, i.e., $I=1$, behave like distinguishable particles, i.e., $\beta = p_B/p_D = 1$. Moreover, it is clear that any result between 1 and 2 is also attainable by a specific value of the scalar product $I$.

Furthermore, Eq.24 also suggests that despite the fact that the bunching parameter $\beta$ is measured on a *single* state (it does not matter which, because $\beta$ is a constant) it carries information $I$ about the overlapping between *all the other states*. Therefore, the states orthogonality can be evaluated experimentally by measuring the bunching parameter.

As was stated in the introduction, there is some similarity between the bunching parameter and the second order correlation function, which was used extensively in the literature [19,20]. However, while the bunching parameter is a generic property, and depends exclusively on $I$, the second order correlations function depends on the states $\psi_1$ and $\psi_2$ as well, and therefore cannot be a constant of motion (like $\beta$). The correlation function of different states would have different values, and therefore cannot be used as a universal measure of bunching.

The differences between the second order correlation function and the bunching parameter will be discussed elsewhere.

## 5. Spatial Bunching

The same reasoning applies to spatial measurements, in which case the eigenstates are delta functions and the summation is replaced by an integral[17]. If the initial wavefunctions are $\psi_1(x)$ and $\psi_2(x)$ then the joint initial wavefunction can be written as

$$\Psi(x_1, x_2; t=0) = \frac{1}{\sqrt{2N}}[\psi_1(x_1)\psi_2(x_2) + \psi_1(x_2)\psi_2(x_1)] \qquad (25)$$

With the same normalization constant

$$N \equiv \frac{1}{2}\int\int |\psi_1(x_1)\psi_2(x_2)+\psi_1(x_2)\psi_2(x_1)|^2 dx_1 dx_2 = 1+|I|^2 \tag{26}$$

where in the continuous case

$$I \equiv \langle \psi_1 | \psi_2 \rangle = \int \psi_1(x)\psi_2^*(x)dx. \tag{27}$$

For $t>0$ each one of the single particle wavefunctions experiences Schrödinger dynamics, but the normalization constant remains $N$, i.e.

$$\Psi(x_1,x_2;t>0) = \frac{1}{\sqrt{2N}}[\psi_1(x_1,t)\psi_2(x_2,t)+\psi_1(x_2,t)\psi_2(x_1,t)] \tag{28}$$

Therefore, now we can ask the question about the probability to measure the two bosons *at the same location*) as often discussed in the literature (with the exception of the scenarios, which are described in Ref.[14], and [15]), in which case, this probability reads

$$p_B(x,t) = \frac{2|\psi_1(x,t)\psi_2(x,t)|^2}{1+|I|^2} \tag{29}$$

for any $x$ and for any $t$, while the probability density for distinguishable particles is

$$p_D(x,t) = |\psi_1(x,t)\psi_2(x,t)|^2. \tag{30}$$

Then again the ratio between these two probabilities is a measure of the spatial bunching

$$\beta(x,t) = \frac{p_B(x,t)}{p_D(x,t)} = \frac{2}{1+|I|^2}, \tag{31}$$

This ratio is independent of $x$ and of $t$, i.e. it is a constant throughout space and time, and depends only on the scalar product between the initial states. Again, this is a number between 1 and 2, where 2 is attributed only to initially orthogonal states.

As was mentioned above, there is one exception to Eq.31. When one of the wavefunctions ($\psi_1(x,t)$ or $\psi_2(x,t)$) vanishes at a certain point $x$ then clearly the bunching parameter has a meaning only as a limit (e.g., l'Hospital). This case is covered in Ref. [14] and will be ignored in this paper.

One of the surprising conclusions of (31) is that by measuring the bunching parameter $\beta(x,t)$ at *any point in space and time*, the overlap between the wavefunctions can be evaluated, i.e., *local* measurement carry the information about the *global* profile of the wavefucntions.

This universal property of $\beta$ can help in the prediction of experimental results *without the need for calculations*. In view of this universality, the results of the beam splitter experiments (see, for example, Ref.[21] for photons and Ref.[22] for massive paricles) are straightforward: it is clear that bosons bunch, in a beam splitter experiment, only when they enter different arms. In this case the initial states are orthogonal and therefore the probability of the bosons to reach the same detector is twice the probability of distinguishable particles to get there. Since the probability distinguishable particles is 1/4, then the bosons must go *together* to either detectors with probability 1/2. The probability to separate is zero. When they both enter the same arm no bunching occurs and the bosons behave like distinguishable particles.

## 6. Generalization of the two-states space

Let us generalize the two-state scenario. In this section we focus on the beam-splitter systems, however, all the conclusions can be applied to any two-state systems. Therefore, the "up" and "down" states of the splitter can be read as the "up" and "down" states of a spin or any other two-state system.

The most general case, is when the two bosons have an arbitrary superposition between the upper and the lower states, i.e.,

$$|\psi_1\rangle = \cos(\theta_1)\exp(i\mu_1)|u\rangle + \sin(\theta_1)\exp(-i\mu_1)|d\rangle \tag{32}$$

$$|\psi_2\rangle = \cos(\theta_2)\exp(i\mu_2)|u\rangle + \sin(\theta_2)\exp(-i\mu_2)|d\rangle \tag{33}$$

where $\theta_1$, $\theta_2$, $\mu_1$ and $\mu_2$ are parameters, which determine the initial states, in which case

$$I^2 = |\langle \psi_1 | \psi_2 \rangle|^2 = \cos^2(\theta_1 - \theta_2) - \sin(2\theta_1)\sin(2\theta_2)\sin^2(\mu_1 - \mu_2) \qquad (34)$$

which implies

$$\beta(\theta_1, \theta_2, \mu_1 - \mu_2) = \frac{2}{1 + \cos^2(\theta_1 - \theta_2) - \sin(2\theta_1)\sin(2\theta_2)\sin^2(\mu_1 - \mu_2)}. \qquad (35)$$

The bunching parameter $\beta$ is therefore very sensitive on the specific initial states of the particles. Only when $\theta_1 - \theta_2 = \pm\pi/2$ and $\mu_1 - \mu_2 = 0, \pm\pi$ or that one of the $\theta$'s is equal to $0$ or $\pm\pi$ (which are the orthogonality requirements) does $\beta = 2$. When $\theta_1 = \theta_2$ and $\mu_1 = \mu_2$ then $\beta = 1$. In general it can have any value between 1 and 2.

If the initial state is picked *randomly* then the probability density of measuring any value of $\beta$ is plotted in Fig.4. For any possible combination of the angles $\theta_1$, $\theta_2$, $\mu_1$ and $\mu_2$, which were selected randomly but uniformly in $[0, 2\pi)$ (this is the entire space of the initial state, when there is no preference of one state over the rest), the distribution of the bunching parameter $\beta$ was calculated. In Fig.4 the histogram of this distribution is plotted, i.e., the frequencies of every value of $\beta$. When all the possible input combinations are taken under consideration then $\rho(\beta)$ is the probability density to measure the value $\beta$, i.e. $\int_{\beta_1}^{\beta_2} \rho(\beta')d\beta'$ is the probability to measure $\beta$ between $\beta_1$ and $\beta_2$.

The mean of all these values of $\beta$ can easily be calculated (it is marked with a circle in Fig.4 ) $\langle\beta\rangle = 1.39$ and the standard deviation is $\sigma = 0.3$. Therefore, for arbitrary two bosons that enters the beam splitter the bunching parameter (either at the input or at the output) is

$$\beta = 1.39 \pm 0.3, \qquad (36)$$

which shows that the average value of bunching is closer to 1 than to 2.

Similarly, from Fig.4 one finds that the probability for $p_B/p_D=1$ is considerably higher than the probability for $p_B/p_D=2$, i.e., the probability for non-bunching is higher than the probability for bunching.

Therefore, randomly chosen bosons pair has a bunching parameter, which is closer to 1 than to 2.

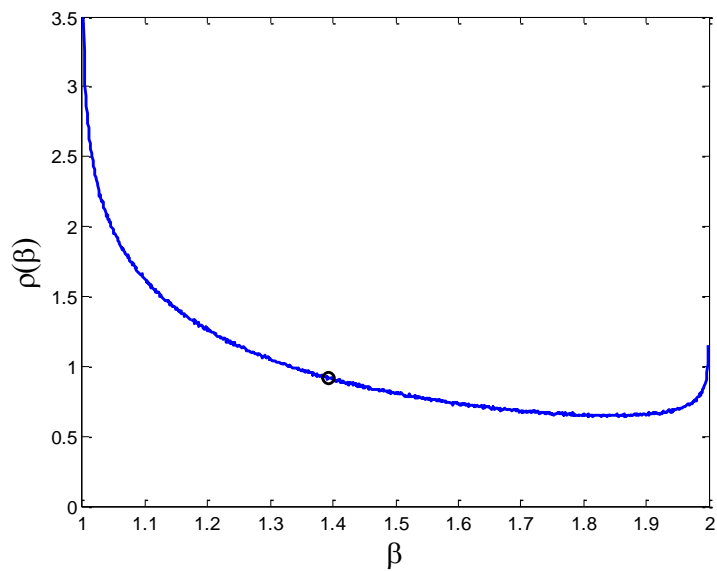

Figure 4: The probability density to measure the values of $\beta = p_B/p_D$ for all possible initial states, which are uniformly determined by the parameters $\theta_1$, $\theta_2$, $\mu_1$ and $\mu_2$. The average value is marked by a circle.

It is also of interest to calculate the bunching parameter when the two initial states have the same phase, in which case the expressions become considerably simpler,

$$\beta(\theta_1 - \theta_2) = \frac{2}{1+\cos^2(\theta_1 - \theta_2)}, \tag{37}$$

for which case

$$\beta = 1.41 \pm 0.34. \tag{38}$$

This result is quite surprising, since despite the increase in the similarity the bunching parameter increases. Moreover, in this case the probability density of the histogram can be calculated analytically

$$\rho(\beta) = \frac{4/\pi}{\sqrt{1-(4/\beta-3)^2 \beta^2}},\qquad(39)$$

and it is presented in Fig.5.

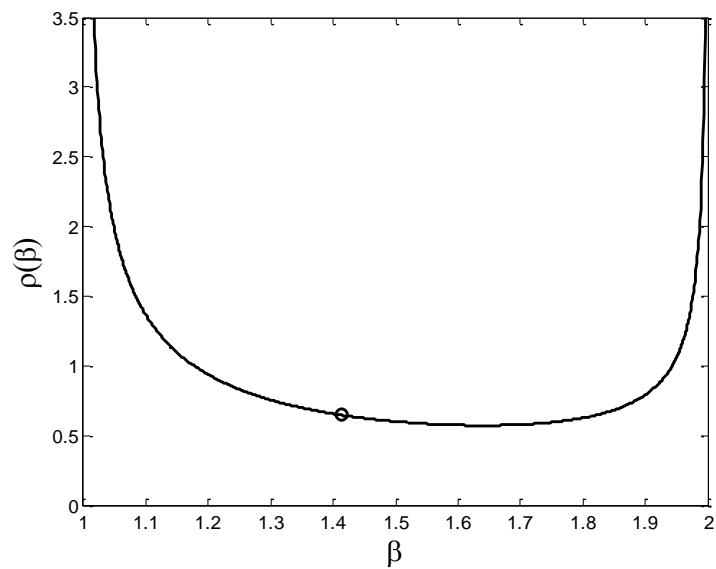

Figure 5: The same as Fig.4 but for all initial states with the same phases, i.e. $\mu_1 = \mu_2$. The mean value ($\sqrt{2}$) is marked by a circle.

Eq. 39 can be approximated at the vicinity of $\beta = 1$ by

$$\rho(\beta \cong 1) = \frac{1}{\pi}\sqrt{\frac{2}{\beta-1}} \qquad(40)$$

and at the vicinity of $\beta = 2$ by

$$\rho(\beta \cong 2) = \frac{1}{\pi\sqrt{2(2-\beta)}}.\qquad(41)$$

Therefore

$$\lim_{\Delta\beta \to 0} \frac{\rho(1+\Delta\beta)}{\rho(2-\Delta\beta)} = 2, \tag{42}$$

which means that the probability to measure a value of $\beta$ close to 1 is twice the probability to measure a value close to 2.

Finally, in the case where $\theta_1 = \theta_2$, i.e., when the amplitudes of the initial states are equal but not necessarily their phases, the measured values of $\beta$ for any randomly chosen state, are much closer to 1, since

$$\beta = \frac{1}{1 - \sin^2(2\theta)\sin^2(\mu_2 - \mu_1)/2}. \tag{43}$$

The probability density of measuring any value of $\beta$ (i.e., the histogram for Eq.43 for all values of $\theta \equiv \theta_1 = \theta_2$ and $\mu_2 - \mu_1$) is plotted in Fig.6.

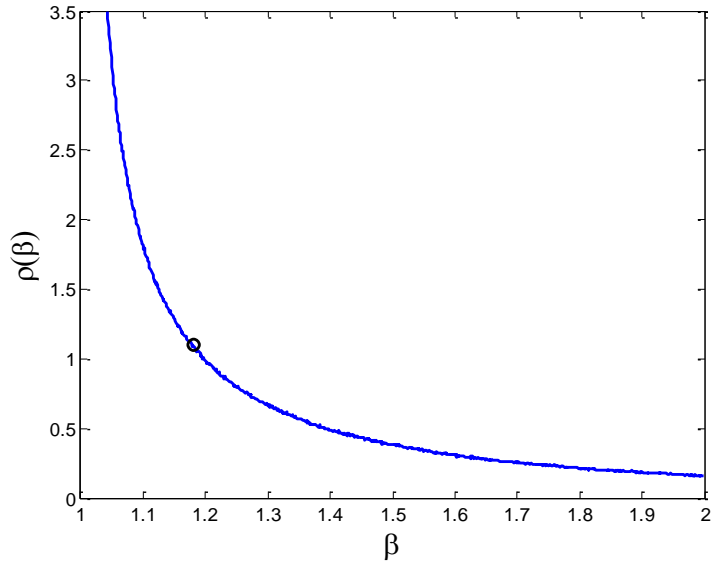

Figure 6: The same as Fig.4 but for all initial states with the same amplitudes, i.e. $\theta_1 = \theta_2$. The mean value ($1.18$) is marked by a circle

In this case the mean value, which is marked by a circle, is much closer to 1:

$$\beta = 1.18 \pm 0.24. \tag{44}$$

## 7. Non-Orthogonality in Continuous space

Assume a scenario, in which two particles propagates towards each other with a relative velocity $v = 2\hbar k/m$, where at $t=0$ the distance between them is $2x_0$ (see Fig.7), then let their wavefunction be described by two Gaussians with a spatial width $\sigma$, i.e.,

$$\psi_1(x) = \frac{1}{\sqrt{\sigma}}\left(\frac{2}{\pi}\right)^{1/4} \exp\left(-\left(\frac{x-x_0}{\sigma}\right)^2 - ikx\right) \qquad (45)$$

and

$$\psi_2(x) = \frac{1}{\sqrt{\sigma}}\left(\frac{2}{\pi}\right)^{1/4} \exp\left(-\left(\frac{x+x_0}{\sigma}\right)^2 + ikx\right) \qquad (46)$$

Respectively.

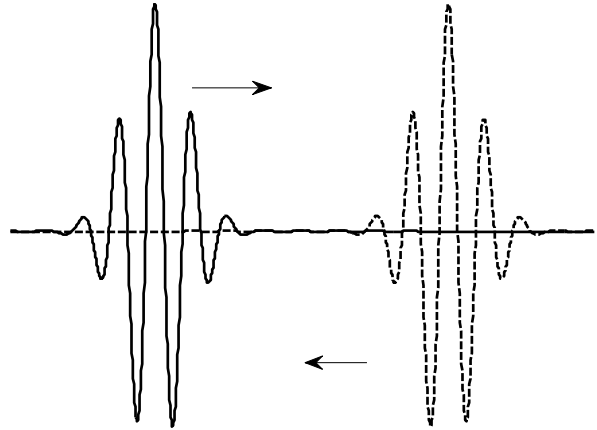

Figure 7: Two Gaussian packets propagating towards each other (in the figure the real part is plotted).

The initial overlap between them is

$$I = \int \psi_1(x)\psi_2^*(x)dx = \exp\left(-2\left(\frac{x_0^2}{\sigma^2}\right) - \frac{k^2\sigma^2}{2}\right). \qquad (47)$$

Despite the fact that the two particles propagates towards each other, and even after approximately a time period $t_0 = m\dfrac{x_0}{\hbar k}$ the two wavepackets overlap, then due to the opposing wavenumber, their overlap integral remains

$$I(x,t) = I(y,0) \text{ for any } x \text{ and } y \tag{48}$$

throughout the dynamics, and the bunching parameter remains

$$\beta = \frac{p_B}{p_D} = \frac{2}{1 + \exp\left(-4\left(\dfrac{x_0^2}{\sigma^2}\right) - k^2\sigma^2\right)}. \tag{49}$$

In most practical cases this value is close to 2, however, if the wavepackets are initially close. For a given distance ($2x_0$) and relative velocity $2\hbar k/m$ between the two particles, the optimum packets width (the one with the highest overlap) is $\sigma = \sqrt{2x_0/k}$, for which case

$$\beta = \frac{p_B}{p_D} = \frac{2}{1 + \exp(-4kx_0)}, \tag{50}$$

which teaches that in order to measure non-bunching conduct the distance between them should not be considerably larger than the mean packet's wavelength.

If, on the other hand, the two particles emerge from *the same* spot, say $x = 0$ and propagate away from each other (see Fig.8), then we can use the same expressions (48) and (49) with $x_0 = 0$ to obtain

$$I = \int \psi_1(x,t)\psi_2^*(x,t)dx = \exp\left(-\frac{k^2\sigma^2}{2}\right) \tag{51}$$

and a constant bunching ratio

$$\beta = \frac{p_B}{p_D} = \frac{2}{1 + \exp(-k^2\sigma^2)}, \tag{52}$$

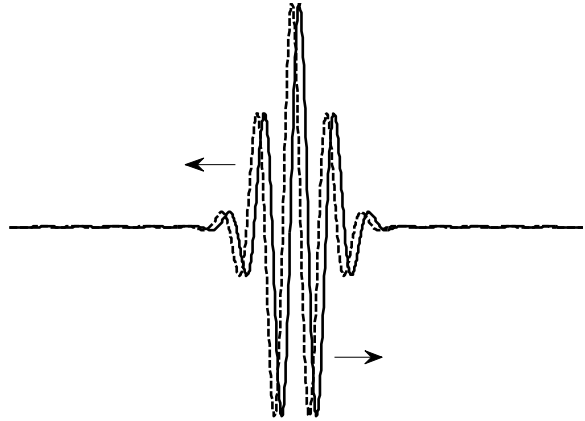

Figure 8: Two Gaussian packets propagating away from each other (in the figure the real part is plotted).

which can, in principle, be arbitrarily close to 1 (when $\sigma \rightarrow 0$).

One of the curious conclusions of this equation (and the direct result of the previous sections) is that by measuring $\beta$ at any point in space and at any time one can calculate the product of the velocity and the initial width $k\sigma$.

It is possible to trap and release heavy bosons in an optical or magnetic traps (see, for example Ref.[23]). In such scenarios the bosons can be 4He, i.e., their mass is 4au, and since the trap dimensions can easily be tens of micrometers, then if the release velocity is lower than 1cm/s then $k\sigma$ can be lower than 1 and $\beta$ lower than 1.5.

## 8. Generalization to $Q$ Particles – Bosonic Enhancement and State Orthogonality

The bunching parameter can be generalized to an arbitrary number of partticles.

For $Q$ bosons, the symmetric wavefunction reads

$$|\Psi\rangle = \frac{1}{\sqrt{Q!N}} \sum_p |p(1):\psi_1; p(2):\psi_2; \cdots p(Q):\psi_Q\rangle \tag{53}$$

where $N$ is the normalization constant,

$$N = \frac{1}{Q!} \sum_{p,q} \prod_{n=1}^{Q} \langle \psi_{p(n)} | \psi_{q(n)} \rangle, \tag{54}$$

and the summation takes over all possible permutations $p$ and $q$.

To calculate the probability to measure all $Q$ bosons in the same final state m, then the state (53) should be projected on the final joint state $|1:m;2:m;\cdots Q:m\rangle$ yielding

$$p_B(m) = \frac{Q!}{N} \prod_{n=1}^{Q} |\langle m | \psi_n \rangle|^2 \tag{55}$$

Similarly, in the case of distinguishable particles

$$p_D(m) = \prod_{n=1}^{Q} |\langle m | \psi_n \rangle|^2 \tag{56}$$

Therefore, any bunching parameter would depend on the orthogonalities between the states via $N$.

Thus, the ratio between the probability to measure all $Q$ bosons at the same spatial point and the probability to measure there all the distinguishable particles is

$$\beta = \frac{p_B(m)}{p_D(m)} = \frac{Q!}{N} \tag{57}$$

Only when all states are orthogonal the well-known result is retrieved. Since then

$$\langle \psi_{p(n)} | \psi_{q(n)} \rangle = \delta[p(n) - q(n)], \tag{58}$$

which yields

$$N = \frac{1}{Q!} \sum_{p,q} \prod_{n=1}^{Q} \delta[p(n) - q(n)] = \frac{1}{Q!} \sum_p 1 = 1, \tag{59}$$

And therefore

$$\beta = Q! \tag{60}$$

In any other scenario,

$$1 \leq \beta < Q! \tag{61}$$

The value $\beta = 1$ is reached when all the particles are initially at the same state.

In the spatial continuous case, the joint wavefunction looks

$$\Psi(x_1, x_2, \ldots x_Q) = \frac{1}{\sqrt{Q!N}} \sum_p \psi_{p(1)}(x_1)\psi_{p(2)}(x_2)\cdots\psi_{p(Q)}(x_Q) = \frac{1}{\sqrt{Q!N}} \sum_p \prod_{m=1}^{Q} \psi_{p(m)}(x_m) \tag{62}$$

where $N$ is the normalization constant,

$$N = \frac{1}{Q!} \sum_{p,q} \prod_{n=1}^{Q} \int \psi_{p(n)}(x)\psi^*_{q(n)}(x)dx = \frac{1}{Q!} \sum_{p,q} \prod_{n=1}^{Q} I[p(n), q(n)] \tag{63}$$

where

$$I[a,b] \equiv \int \psi_a(x)\psi^*_b(x)dx \tag{64}$$

is the overlap integral between states $a$ and $b$, and the summation takes over all possible permutations $p$ and $q$.

Then, the probability density to measure all bosons at the same spatial point $x_0$ is

$$p_B(x_1 = x_2 = \ldots = x_Q = x_0) = \frac{Q!}{N} \prod_{n=1}^{Q} |\psi_n(x_0)|^2 \tag{65}$$

While for distinguishable particles

$$p_D(x_1 = x_2 = \ldots = x_Q = x_0) = \prod_{n=1}^{Q} |\psi_n(x_0)|^2. \tag{66}$$

and thus the bunching parameter is again

$$\beta = \frac{p_B(x_0, x_0, \ldots x_0)}{p_D(x_0, x_0, \ldots x_0)} = \frac{Q!}{N} = \frac{(Q!)^2}{\sum_{p,q} \prod_{n=1}^{Q} I[p(n), q(n)]} \qquad (67)$$

The well-known bosonic enhancement factor $\beta = Q!$, which appears in multiple bosons systems[24-27], is a consequence of the states orthogonality. In any other case, where the overlapping integrals between states do not vanish, the bunching parameter is smaller.

## 9. Application to Other Degrees of Freedom

Despite the fact that we focused on spinless bosons, the entire treatment, and the definition of the bunching parameter or the enhancement factor can be applied to any degree of freedom including the spin. For example, if the initial states are

$$\left|\psi_1^{in}\right\rangle = \cos(\alpha_1) e^{i\delta_1} |u\rangle + \sin(\alpha_1) e^{-i\delta_1} |d\rangle, \qquad (68)$$

$$\left|\psi_2^{in}\right\rangle = \cos(\alpha_2) e^{i\delta_2} |u\rangle + \sin(\alpha_2) e^{-i\delta_2} |d\rangle \qquad (69)$$

where $|u\rangle$ and $|d\rangle$ stand for the two orthogonal spin's states,

then still the bunching parameter or the enhancement factor can be defined and measured as

$$\beta = \frac{2}{1 + \left|\left\langle \psi_1^{in} | \psi_2^{in} \right\rangle\right|^2} = \frac{2}{1 + \left|\cos(\alpha_1)\cos(\alpha_2) e^{i(\delta_2 - \delta_1)} + \sin(\alpha_1)\sin(\alpha_2) e^{i(\delta_1 - \delta_2)}\right|^2} . \qquad (70)$$

## 10. Summary of the Main Conclusions

It is well-known that the bunching parameter $\beta \equiv p_B / p_D$ of bosons was calculated and measured to be equal to 2. This phenomenon was termed bosons bunching as well as bosons enhancement. However, that is because in all these cases the initial particles' states were mutually orthogonal. In this paper the generic case is solved without the orthogonality requirement. It is shown that $\beta(x,t) = \beta$ is a constant in space and time and is equal to $\beta = 2/(1+I^2)$ where $I \equiv \langle \psi_1 | \psi_2 \rangle$ is the overlap between the initial states. Therefore, if the initial states are not orthogonal, i.e., $|I| > 0$, then $\beta$ can be lower than 2, and if the initial states are identical then $\beta = 1$ and no bosons bunching occurs. This result was generalized to an arbitrary number of particles $Q$, in which case the bunching parameter $\beta = (Q!)^2 / \sum_{p,q} \prod_{n=1}^{Q} I[p(n), q(n)]$ is equal to the well-known bosonic enhancement factor $Q!$ only when all the states are orthogonal.

It should be stressed that since the symmetrization postulate is universal, the results presented in this papers are valid for any Hamiltonian, for any degree of freedom and for any interaction.

This universality can help in the prediction of experimental results without the need for calculations. For example, using the dependence of $\beta$ on $I$ it is straightforward to predict *without taking any calculations* that bosons bunch, in a beam splitter experiment, only when they enter different arms (as was explained in the previous section). When they both enter the same arm no bunching occurs and the bosons behave like distinguishbale particles.